\documentclass[twocolumn,showpacs,preprintnumbers,amsmath,amssymb]{revtex4}
\usepackage{graphicx}% Include figure files
\usepackage{dcolumn}% Align table columns on decimal point
\usepackage{bm}% bold math

\begin{document}

\preprint{}

\title{Optical properties and electronic structure of ZrB$_{12}$}% Force line breaks with \\

\author{J. Teyssier}
\author{A. B. Kuzmenko}
\author{D. van der Marel}
\author{F. Marsiglio}

\affiliation{D\'epartement de Physique de la Mati\`ere
Condens\'ee, Universit\'e de Gen\`eve, Quai Ernest-Ansermet 24,
1211 Gen\`eve 4, Switzerland}%Lines break automatically or can be forced with \\

\author{A. B. Liashchenko}
\author{N. Shitsevalova}
\author{V. Filippov}

\affiliation{Institute for Problems of Materials Science, National
Academy of Sciences of Ukraine, 252680 Kiev, Ukraine}
\date{\today}

\begin{abstract}
We report optical (6 meV - 4 eV) properties of a boride
superconductor ZrB$_{12}$ ($T_c$ = 6 K) in the normal state from
20 to 300 K measured on high-quality single crystals by a
combination of reflectivity and ellipsometry. The Drude plasma
frequency and interband optical conductivity calculated by
self-consistent full-potential LMTO method agree well with
experimental data. The Eliashberg function $\alpha_{tr}^2F(\omega)$
extracted from optical spectra features two peaks at about 25 and
80 meV, in agreement with specific heat data. The total coupling constant is $\lambda_{tr}=1.0\pm0.35$. The low
energy peak presumably corresponds to the displacement mode of Zr
inside $B_{24}$ cages, while the second one involves largely boron
atoms. In addition to the usual narrowing of the Drude peak with
cooling down, we observe an unexpected removal of about 10 \% of
the Drude spectral weight which is partially transferred to the
region of the lowest-energy interband transition ($\sim$ 1 eV).
This effect may be caused by the delocalization of the metal ion
from the center of the $B_{24}$ cluster.
\end{abstract}

\maketitle

\section{Introduction}

Boron, like carbon, covalently bonds to itself, easily forming rigid
three-dimensional clusters and networks as well as planes, chains
and even nanotubes. The discovery of superconductivity in
graphite-like MgB$_{2}$ at 40 K \cite{Nagamatsu2001} has stimulated
intense research of other superconducting boron phases, although
they have so far shown quite modest transition temperatures.
Zirconium dodecaboride ZrB$_{12}$ has one of the highest known
$T_{c} \approx $ 6 K among the binary borides except MgB$_{2}$. Even
though the superconductivity there has been discovered almost 40
years ago \cite{Matthias1968}, only recent progress in the
single-crystal growth enabled its extensive studies.

In dodecaborides MB$_{12}$, boron atoms make up a three-dimensional
network forming spacious B$_{24}$ cages which accommodate metal
ions. The isotope effect in ZrB$_{12}$ for zirconium
\cite{ChuScience68} ($\beta\approx$ -0.32) is much larger than for
boron \cite{FiskPL71} ($\beta\approx$ -0.09), pointing to a large
contribution of lattice modes involving Zr atoms to the
electron-phonon coupling responsible of superconductivity in ZrB$_{12}$.
The inversion of specific heat measurements \cite{Lortz2005} gives a
pronounced peak in the phonon density of states at about 15 meV,
attributed to Zr vibrations in oversized boron cages. The coupling
to this mode can also be enhanced by strong anharmonicity. The
contribution to electron phonon coupling of this mode was also
reported in Seebeck effect measurements \cite{Glushkov2006}.

Notably, significant controversy exists regarding the strength of
the electron-phonon interaction. According to the McMillan formula,
a weak-to-medium coupling $\lambda\approx 0.68$ provides the
observed $T_{c}$, given the phonon mode frequency of 15 meV and a
'standard' value of the screened Coulomb potential
$\mu^{\ast}\approx 0.1-0.15$ \cite{Daghero2004}. A weak coupling was
confirmed by specific heat measurements \cite{Lortz2005}. However,
tunneling \cite{Tsindlekht2004} and point-contact spectroscopy
\cite{Daghero2004} yield $2\Delta_{0}/k_{B} T_{c}\approx 4.8$
suggesting a very strong coupling regime. This discrepancy between
the results of surface and bulk probes has been ascribed to the
surface-enhanced superconductivity in ZrB$_{12}$
\cite{Tsindlekht2004,Khasanov2005}. In principle, the value of
$\lambda$ can be also derived from the DC resistivity measurements,
although this calculation relies on the unknown value of the plasma
frequency.

Band structure calculation \cite{Shein2003} shows that B 2p and Zr 4d
states contribute almost equally to the density of states at the
Fermi level. This makes ZrB$_{12}$ electronically very different
from MgB$_{2}$, where Mg states have a vanishing contribution to the
metallic and superconducting properties \cite{Kortus2001}. However,
no experiments which can directly verify the prediction of the band
theory, such as angle-resolved photoemission (ARPES), de Haas-van
Alphen (dHvA) effect, have been done so far on this compound.

Optical spectroscopy that probes free carrier charge dynamics and
provides experimental access to the plasma frequency, electron
scattering and interband transitions, may help to clarify a number
of existing uncertainties and test the predictions of the band
calculations. By a combination of reflectivity and ellipsometry, we
have obtained spectra of optical conductivity and dielectric
function in the broad range of photon energies and frequencies. We
concentrate on general optical fingerprints of the electronic
structure of ZrB$_{12}$ in the normal state.

For comparison, we performed calculations of the band structure and
corresponding optical properties. The electron-phonon coupling
function $\alpha_{tr}^2F(\omega)$ has been extracted from the fit of
optical data and compared with thermodynamic and Raman measurements.

An unusual decrease of the Drude plasma frequency with cooling down
is observed. By comparing with other clustered compounds, we found
that the delocalization of the metal ion (weakly bonded to the boron
network) from the center of the boron cage could be at the origin of
large changes in the electronic properties.

\section{Crystal Structure}

\begin{figure}
\includegraphics[width=6cm]{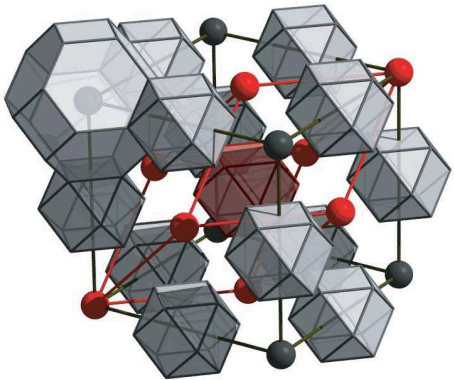}\\
\caption{(color on-line) Crystal structure of ZrB$_{12}$. The
clusters B$_{12}$ and B$_{24}$ are displayed as polyhedrons. The
unit cell containing 13 atoms is colored in red.}\label{structure}
\end{figure}

ZrB$_{12}$ crystallizes in the UB$_{12}$ type structure
\cite{Leithe2002} which can be viewed as a cubic rocksalt
arrangement of Zr and B$_{12}$ cuboctahedral clusters
(Fig.\ref{structure}). For our discussion, it can be more convenient
to represent this structure as a face centered cubic structure of
Zr, surrounded by $B_{24}$ cages.

\section{Optical experiment and results}

Large single crystals of ZrB$_{12}$ were grown using a high
frequency induction zone furnace \cite{Paderno2002}. Optical
measurements were performed on (001) surface with dimensions $4
\times 4$ mm$^2$. The surface of this extremely hard material was
polished using diamond abrasive disks with finest grain size of
$0.1$ $\mu$m. In the photon energy range $0.8$-$4$ eV, the complex
dielectric function $\epsilon(\omega)$ was determined directly using
spectroscopic ellipsometry at an incident angle of $60^{\circ}$. For
photon energies between $6$ meV and $0.8$ eV, the reflectivity of
the sample was measured using a Bruker 113 Fourier transform
infrared spectrometer. For both ellipsometry and reflectivity
experiments, the sample was mounted in a helium flow cryostat
allowing measurements from room temperature down to 7 K. The
reference was taken by \emph{in situ} gold evaporation.

\begin{figure}[htp]
\includegraphics[width=8.5cm]{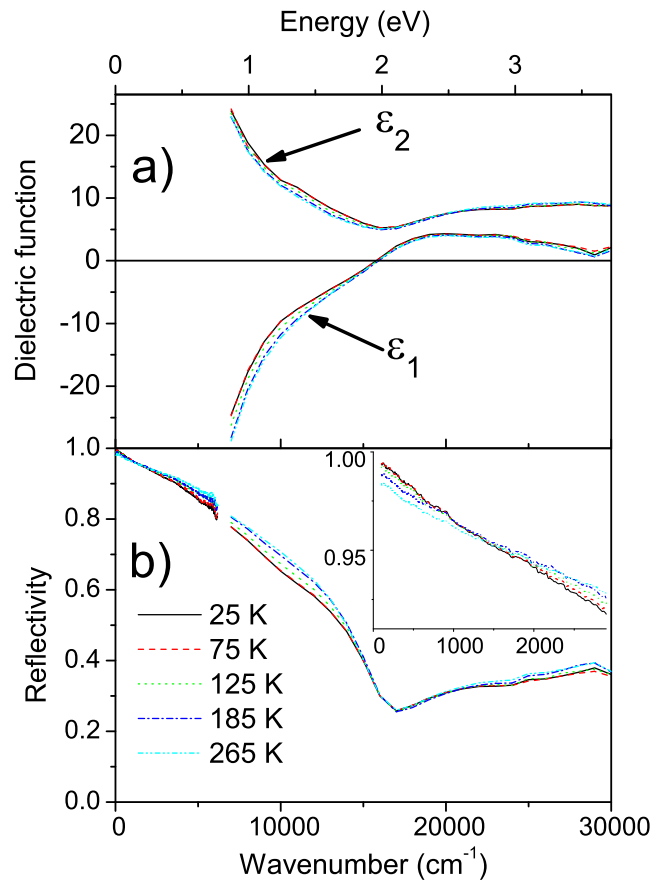}\\
\caption{(color on-line) a) Real and imaginary part of the
dielectric function measured by ellipsometry. b) Reflectivity
measured in the infrared and calculated from the dielectric constant
at higher frequencies. Inset: low frequency region.} \label{opt_exp}
\end{figure}

Fig.\ref{opt_exp}a presents the real and imaginary parts of
$\epsilon(\omega)$ measured by ellipsometry at selected
temperatures. The reflectivity measured at low frequencies and
extracted from the dielectric constant in the visible range, are
plotted in Fig.\ref{opt_exp}b. 

In order to obtain the optical conductivity in the infrared region
we used a variational routine \cite{Kuzmenko2005} yielding the
Kramers-Kronig consistent dielectric function which reproduces all
the fine details of the infrared reflectivity data while {\em
simultaneously} fitting to the complex dielectric function in the
visible and UV-range. This procedure anchors the phase of the
infrared reflectivity to the phase at high energies measured with
ellipsometry \cite{bozovic1990}.

\begin{figure}
\includegraphics[width=8.5cm]{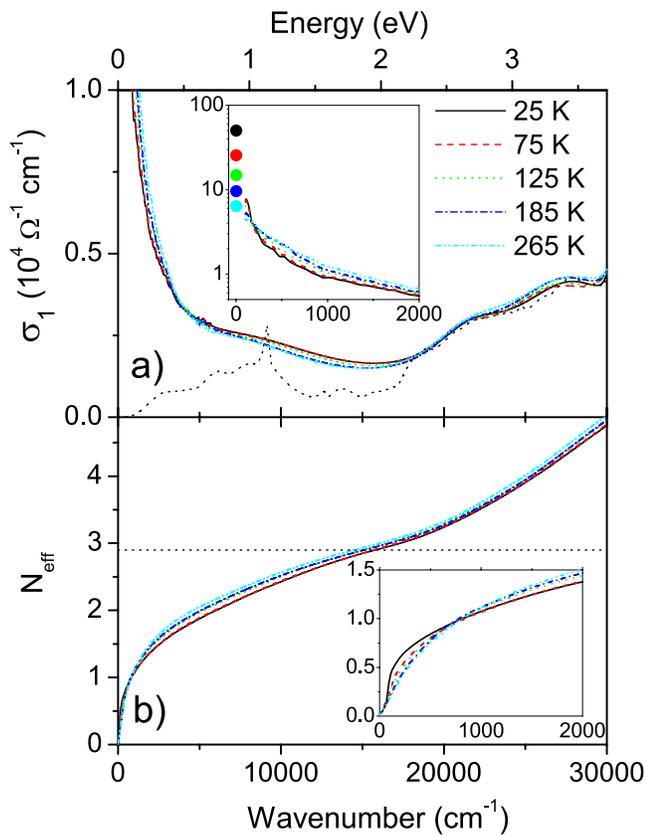}\\
\caption{(color on-line) a) Optical conductivity of ZrB$_{12}$
 at selected temperatures. Inset: low energy part, symbols
represent $\sigma_{\mbox{\tiny DC}}$ from \cite{Daghero2004}.The
dotted line shows the interband conductivity from LDA calculation.
b) Effective number of carriers per unit cell of 13 atoms, inset:
low frequency part. The horizontal dotted line corresponds to the
calculated plasma frequency of 6.3 eV.}\label{S1temp}
\end{figure}

Fig. \ref{S1temp}a shows the evolution of the optical conductivity
with temperature. One can see a Drude-like peak that indicates a
metallic behavior. The DC conductivity ($\sigma_{\mbox{\tiny{DC}}}$)
\cite{Daghero2004,Lortz2005} plotted as symbols in the inset of
Fig.\ref{S1temp}a, agrees reasonably well with the extrapolation  of
the optical data to zero frequency.

Fig.\ref{S1temp}b depicts the effective number of carriers
\begin{equation}
\label{swf} N_{\mbox{\scriptsize eff}}(\omega)=\frac{2mV_c}{\pi
e^2}\int_0^{\omega}\sigma_1(\omega')d\omega'
\end{equation}
\noindent where $m$ is the free electron mass, $e$ is the electron
charge, $V_c$=101.5 {\AA}$^3$ is the unit cell volume. We observe a
transfer of spectral weight to low energy, which comes from the
narrowing of the Drude peak. For higher energies (above 100 meV) an
unusual decrease of up to 10\% is observed. This will be discussed
in section \ref{sectionSW}.

\section{LDA calculations}
\subsection{Band structure}

\begin{figure}[htp]
\includegraphics[width=8.5cm]{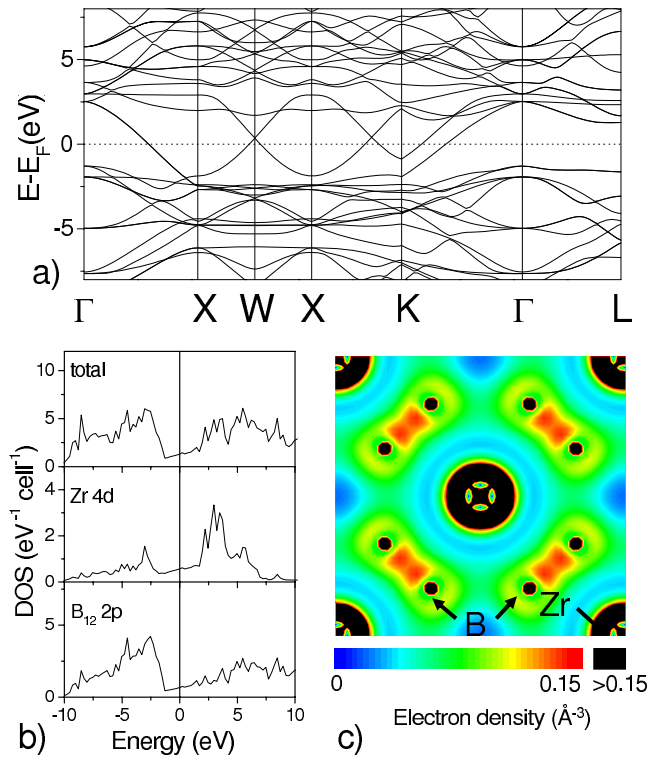}\\
\caption{(color on-line) a) Calculated band structure, b) partial
density of states (PDOS) of ZrB$_{12}$. c) Charge density of in the
a-b plane of the cubic cell.} \label{bands}
\end{figure}

The band structure was calculated for the unit cell containing 13
atoms. We used a full potential plane waves Linear Muffin-Tin
Orbital (LMTO) program \cite{Savrasov1996} within a Local Spin
Density Approximation (LSDA) and Generalized Gradient Approximation
(GGA) \cite{Perdew1996}. The band structure for high symmetry
directions in the Brillouin Zone (BZ) is shown in Fig.\ref{bands}a.
The density of states (DOS) calculated by the tetrahedron method is
displayed in Fig.\ref{bands}b. It appears clearly that the main
contribution to the DOS at the Fermi level comes from zirconium 4d
and boron 2p states. The total DOS at the Fermi level $N(E_F)=1.59$
eV$^{-1}$cell$^{-1}$. $N(E_F)$ for zirconium d electron is 0.55
eV$^{-1}$cell$^{-1}$ and the one for p electrons of B$_{12}$
clusters is 0.65 eV$^{-1}$cell$^{-1}$ which is in good agreement with a
previous report \cite{Shein2003}.

Mapping the repartition of charges in a crystal is helpful to
understand the chemical bonding. The calculated charge density of
valence electrons in the a-b plane of the cubic cell is displayed in
Fig.\ref{bands}c. The high density between direct neighbor boron
atoms indicates a strong covalent bonding while the ion in the
middle of the B$_{24}$ cage (Zr in our case) is weakly bonded to the
boron skeleton. This has strong consequences on vibration modes in
the crystal and may explain the presence of a very low frequency
phonon ($\sim 15$ meV) associated with a displacement of the
zirconium atom in the B$_{24}$ cage.

\begin{figure}[htp]
\includegraphics[width=8.5cm]{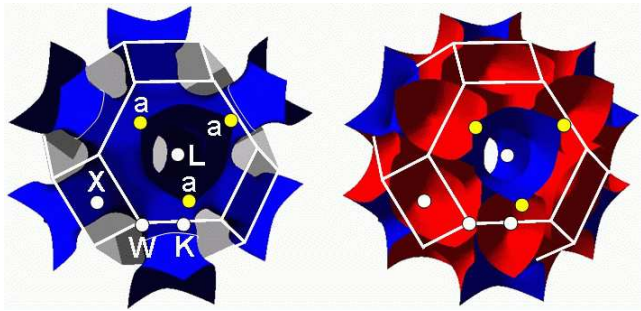}\\
\caption{(color on-line) Fermi surface of ZrB$_{12}$ containing two
sheets: a hole-like network (blue) and electron-like closed surfaces
(red). In the left figure only the first sheet is displayed for
clarity. The points of tangential contact between the two sheets are
designated by symbol $a$.} \label{fermi}
\end{figure}

Fig.\ref{fermi}a shows the Fermi surface in the first Brillouin
zone. The sheet centered at X (marked in red in Fig.\ref{fermi}) has
electron character while the network centered on the $\Gamma$ point
(marked in blue in Fig.\ref{fermi}) is of hole character. The two
sheets are touching at point $a$ along the $\Gamma$ $X$ direction.
In contrast to MgB$_2$, which exhibits a strongly anisotropic Fermi
surface composed of different sheets with very different character
of carriers, the Fermi surface in ZrB$_{12}$ is much more isotropic.

\subsection{Optical properties}

The LDA band structure allows us to calculate the optical
parameters. The interband contribution to the real part of the
optical conductivity can be deduced from the equation:
\begin{equation}
\nonumber\sigma_1(\omega)=\frac{e^2}{12\pi^2m^2\omega}\sum_{f,i}\int_{BZ}
d^3k|P_{fi}|^2\delta [E_f(k)-E_i(k)-\hbar\omega]
\end{equation}
\begin{equation}
\nonumber P_{fi}=\frac{\hbar}{i}<f|\nabla|i>
\end{equation}
Here, $E_i(k)$ and $E_f(k)$ are the energies of the initial
(occupied) and final (empty) states, respectively. $k$ is the wave
vector inside the BZ where the transition $E_i(k)\rightarrow E_f(k)$
occurs. Only direct transitions are taken into account. The
calculated optical conductivity is shown as the dotted line in
Fig.\ref{S1temp}a together with experimental curves. At frequencies
above 2 eV a very good agreement is observed. At lower frequencies,
the experimental curves are systematically higher likely due to the
tail of the Drude peak.

The plasma frequency was computed through integration on the Fermi
surface :
\begin{equation}\label{plasma}
\Omega_{p,LDA}^2=\frac{e^2}{12 \pi^2}\int_{FS}v_fdS
\end{equation}
where $v_f$ is the Fermi velocity. We found a value $\Omega_{p,LDA}=6.3$
eV which corresponds to $N_{eff}\approx$ 2.9 (dotted line in
Fig.\ref{S1temp}b). The experimental value of $N_{eff}(\omega)$
reaches this level at about 2 eV where the experimental optical
conductivity (Fig.\ref{S1temp}a) shows a minimum. This indicates
that the calculated plasma frequency is close to the experimental
one.

\section{Analysis and discussion}

\subsection{Extended drude analysis}

\begin{figure}
\includegraphics[width=8.5cm]{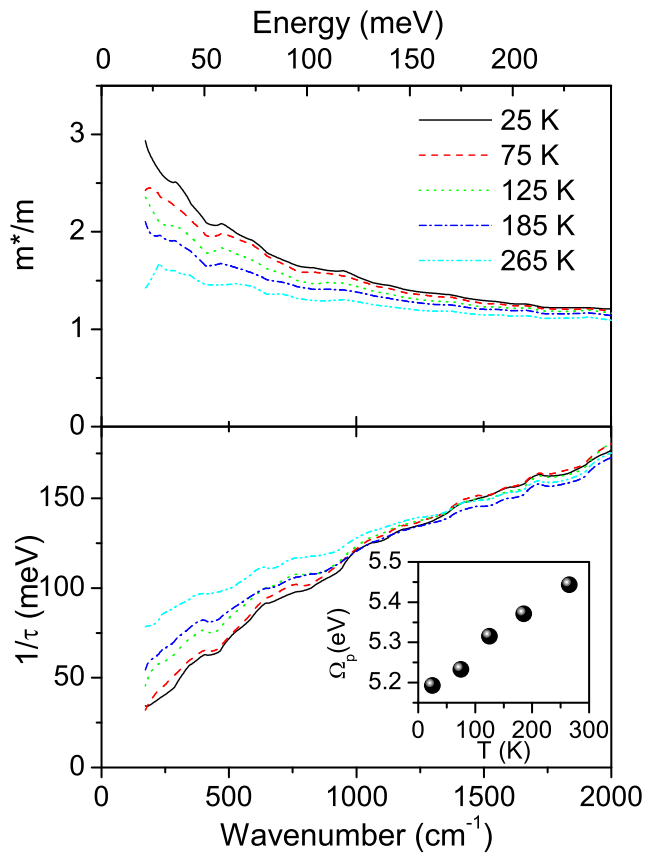}\\
\caption{(color on-line) Extended Drude analysis of the optical
conductivity of ZrB$_{12}$ for different temperatures. Inset shows
the T dependence of the plasma frequency used to calculate $m^*/m$
ans $1/\tau$.}\label{ExDrude}
\end{figure}

Fig.\ref{ExDrude} shows the frequency dependence of the mass
renormalization $m^*(\omega)/m_b$ ($m_b$ is the band mass) and the
scattering rate $1/\tau(\omega)$ obtained by the extended Drude
formalism:
\begin{eqnarray}
\frac{m^*(\omega)}{m_b}&=&-\frac{\Omega_p^2}{4\pi\omega}\mbox{Im}\left(\frac{1}{\sigma(\omega)}\right)\\
\frac{1}{\tau(\omega)}&=&\frac{\Omega_p^2}{4\pi}\mbox{Re}\left(\frac{1}{\sigma(\omega)}\right)
\end{eqnarray}
The plasma frequency $\Omega_p$ was obtained at each temperature by
intergation of optical conductivity from zero to 0.74 eV (6000
cm$^{-1}$) (inset in Fig.\ref{ExDrude}). This cutoff frequency was
chosen such that $\Omega_p$ is close to the plasma frequency
obtained using the fitting procedure (fit 3) described below. The
temperature dependence of $\Omega_p$ is too large to be neglected as
it is usually done.

At low frequencies, the decrease of the scattering rate with cooling
down indicates a narrowing of the Drude peak. The strong frequency
dependence below 100 meV of $1/\tau(\omega)$ and $m^*(\omega)/m_b$
indicates a significant electron-boson, presumably electron-phonon
interaction.

\subsection{Electron-phonon coupling}

The signature of electron-phonon interaction shown by the extended
Drude analysis makes the simple Drude model inapplicable to describe
the low frequency region. Therefore, we adopted the following model
for the dielectric function:
\begin{equation}\label{drude2}
\epsilon(\omega)=\epsilon_{\infty}-\frac{\Omega_{p}^2}{\omega[\omega+iM(\omega,T)]}+\sum_j
\frac{\Omega_{p,j}^2}{\omega_{0,j}^2-\omega^2-i\omega \gamma_j}
\end{equation}
$\epsilon_{\infty}$ represents the contribution of core electrons,
the second and the third terms describe free carriers and interband
contributions respectively. The latter is taken to be a sum of
Lorentzians with adjustable parameters. The frequency dependent
scattering of the free carriers is expressed via the memory
function:
\begin{equation}\label{M}
M(\omega,T)=\gamma_{imp}-2i \int_{0}^{\infty}d\Omega
\alpha^2_{tr}F(\Omega)K\left(\frac{\omega}{2\pi
T},\frac{\Omega}{2\pi T}\right)
\end{equation}
where $\gamma_{imp}$ is the impurity scattering rate and
$\alpha^2_{tr}F(\Omega)$ is the transport Eliashberg function. The
kernel $K$ is \cite{Dolgov1995}:
\begin{eqnarray}
K(x,y)=&\frac{i}{y}+\left\{\frac{y-x}{x}[\Psi(1-ix+iy)-\Psi(1+iy)]
\right\}-\nonumber\\
&\left\{y\rightarrow-y\right\}
\end{eqnarray}
where $\Psi(x)$ is the digamma function.

The reflectivity at normal incidence is:
\begin{equation}\label{R}
R(\omega)=\left|\frac{\sqrt{\epsilon(\omega)}-1}{\sqrt{\epsilon(\omega)}+1}\right|^2
\end{equation}
We fit simultaneously $R(\omega)$ at low frequencies and both
$\epsilon_1(\omega)$ and $\epsilon_2(\omega)$ at high frequencies.

\begin{figure}
\includegraphics[width=8cm]{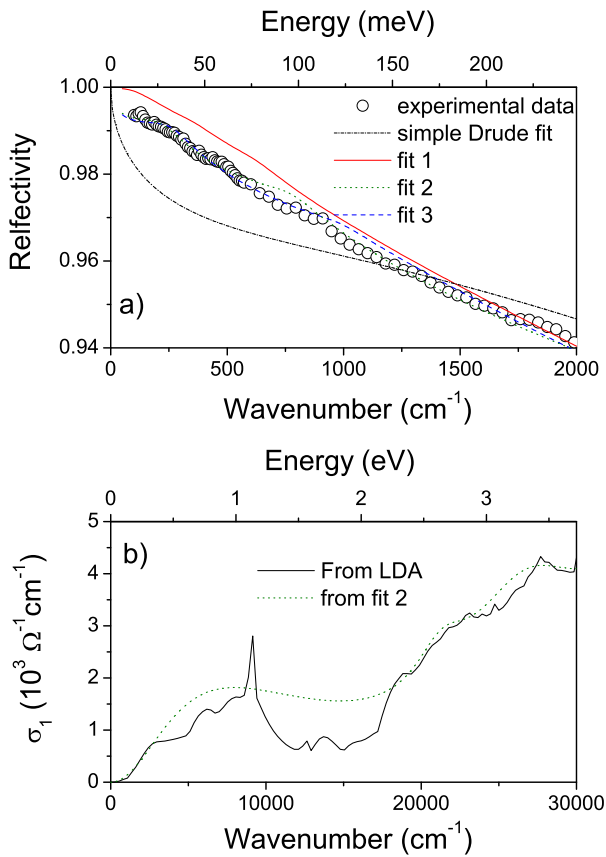}\\
\caption{a) Different fits of reflectivity at 25K. dash-dotted line
is a fit with a single Drude peak. Fit 1 with $\alpha^2F(\Omega)$
from \cite{Lortz2005} (solid red curve). Fit 2 and 3
$\alpha_{tr}^2F(\Omega)$ was adjustable as described in the text.
b)Interband contribution to optical conductivity calculated from LDA
(solid curve) and obtained from the Fit 3 as described in the
text.}\label{R_fit}
\end{figure}

\begin{figure}
\includegraphics[width=8cm]{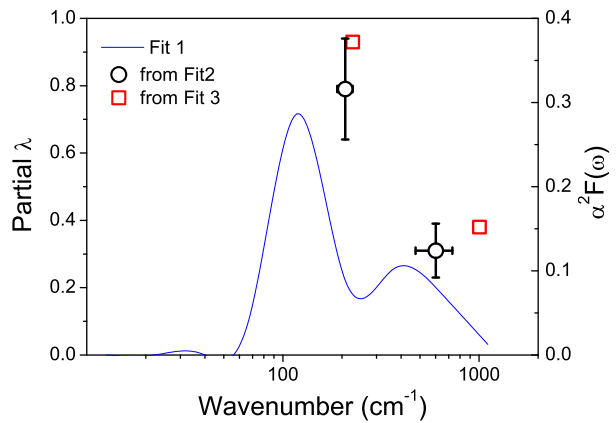}\\
\caption{(color online) The solid curve (fit 1) corresponds to
$\alpha^2F(\Omega)$ from \cite{Lortz2005}. Central phonon
frequencies and partial coupling constant extracted from fit of
optical data with fit 2 (cicles) and fit 3 (squares) procedures as
described in the text.}\label{a2f_fig}
\end{figure}

As a starting point, we used $\alpha^2F(\Omega)$ deduced from
specific heat and resistivity measurements (neglecting the difference between
isotropic $\alpha^2F(\Omega)$ and transport
$\alpha^2_{tr}F(\Omega)$) \cite{Lortz2005} and adjusted all other
parameters (fit 1: solid red curve in Fig.\ref{R_fit}a). This
function (solid curve in Fig.\ref{a2f_fig}) has two peaks around 12
meV and 60 meV and gives $\lambda=0.68$. A clear improvement of the fit quality was achieved
compared to one with a single Drude peak (dashed dotted curve) with
frequency independent scattering rate. The reflectivity curve is within absolute reflectivity error bars of 0.5\%. This means that optical data are consistent with the mentioned value of the coupling constant and the observed $T_c$ \cite{Lortz2005,Daghero2004}.\\

It is also important to determine the spread of parameters consistent with optical data alone. It is known that the extraction of $\alpha^2F(\Omega)$ from optical data is an ill-posed problem \cite{Marsiglio1999,Shulga2001,Dordevic2005}which means that fine details cannot be extracted. We model the
Eliashberg function as a superposition of Dirac peaks:
\begin{equation}\label{a2F}
\alpha^2_{tr}F(\Omega)=\sum_k A_k \delta(\Omega-x_k)
\end{equation}
and treat $A_k$ and $x_k$ as free parameters. We did not succeed to obtain a reasonable fit with a single mode even using some gaussian broadening within 10 meV but two peaks are sufficient to reproduce the shape of experimental curves. The
central frequencies where found to be $x_1$=208 $\pm$ 19 cm$^{-1}$
and $x_2$=602 $\pm$ 128 cm$^{-1}$. The weights are $A_1$=83 $\pm$ 21
cm$^{-1}$ and $A_2$=94 $\pm$ 35 cm$^{-1}$ respectively. The
electron-phonon coupling constant can be derived from the parameters
of the fit by the relation:
\begin{equation}\label{coup}
\lambda_{tr}=2\int_0^\infty\frac{d\Omega\alpha^2_{tr}F(\Omega)}{\Omega}=2\sum_k\frac{A_k}{x_k}
\end{equation}
In our case, this ratio leads to a partial coupling constant
$\lambda_{tr,1}$=0.82 $\pm$ 0.28 for the mode at $x_1$ and
$\lambda_{tr,2}$=0.18 $\pm$ 0.10 for the mode at $x_2$. The total
$\lambda_{tr}$=1.00 $\pm$ 0.35. The other parameters of the fit are
displayed in table I. The reflectivity corresponding to this fit
(fit 2) is displayed as the dotted (green) curve in Fig.\ref{R_fit}a.

\begin{table}
\label{param_fit} \caption{Set of parameters used in fit 2 and fit 3. In fit 3, the Lorentzian $L_1$ is split into three different ones.}
\begin{tabular}{cc|cccccc}
\hline
\hline
&&\multicolumn{3}{c}{Fit 2}&\multicolumn{3}{c}{Fit 3}\\

\hline
\hline
\multicolumn{2}{c|}{$\epsilon_\infty$}&\multicolumn{3}{c}{3.33}&\multicolumn{3}{c}{3.28}\\
\multicolumn{2}{c|}{$\Omega_p$ (eV)}&\multicolumn{3}{c}{5.5}&\multicolumn{3}{c}{5.2}\\
\multicolumn{2}{c|}{$\gamma_{imp}$ (meV)}&\multicolumn{3}{c}{33}&\multicolumn{3}{c}{28}\\
&&\multicolumn{3}{c}{}&$L_{1a}$&$L_{1b}$&$L_{1c}$\\
\hline
&$\omega_o$ (eV)&\space&1&\space&0.32&0.74&1.21\\
Low Energy ($L_1$)&$\Omega_p$ (eV)&&4.6&&1.03&0.50&5.76\\
&$\gamma$ (eV)&&1.8&&0.30&0.15&2.75\\
\hline
&$\omega_o$ (eV)&&2.6&&\multicolumn{3}{c}{2.8}\\
$L_2$&$\Omega_p$ (eV)&&1.9&&\multicolumn{3}{c}{2.1}\\
&$\gamma$ (eV)&&0.6&&\multicolumn{3}{c}{0.5}\\
\hline
&$\omega_o$ (eV)&&3.3&&\multicolumn{3}{c}{3.4}\\
$L_3$&$\Omega_p$ (eV)&&4.7&&\multicolumn{3}{c}{4.1}\\
&$\gamma$ (eV)&&1.3&&\multicolumn{3}{c}{1.0}\\
\hline
&$\omega_o$ (eV)&&4.8&&\multicolumn{3}{c}{4.7}\\
$L_4$&$\Omega_p$ (eV)&&8.3&&\multicolumn{3}{c}{7.9}\\
&$\gamma$ (eV)&&1.9&&\multicolumn{3}{c}{1.6}\\
\hline
\hline
\end{tabular}

\end{table}

The low frequency mode can be logically assigned to the vibration of
weakly bounded zirconium ion inside B$_{24}$ cages, although the
ab-initio calculation of the phonon structure is needed to
substantiate this assignment. The high frequency mode presumably
corresponds to a rigid boron network vibration; a situation similar
to intercalated fullerenes and other clustered covalent structures.

The error bars of $x_k$ and $A_k$ are quite large because of a
significant numerical fit correlation between the lowest energy
interband transition and the parameters of the oscillator (L1 in
table I) and those of $\alpha^2_{tr}F(\Omega)$. In other words, it
is difficult to decouple the effects of low energy interband
transition and electron-phonon interaction on the optical
conductivity. Therefore we performed another fit (fit 3) where the
interband contribution was modeled (parameters in table I) to match the one of the
interband optical conductivity calculated by LDA (solid line in
Fig.\ref{R_fit}b). The presence of sharp structures slightly worsens
the quality of the fit. The low energy mode of $\alpha_{tr}^2F(\Omega)$
is almost unchanged by this procedure ($x_1$=227 cm$^{-1}$,
$A_1$=106 cm$^{-1}$ and $\lambda_{tr1}$=0.93) but the second mode
tends to shift to higher frequencies ($x_2$=1003 cm$^{-1}$,
$A_2$=190 cm$^{-1}$ and $\lambda_{tr2}$=0.38) and the total coupling
constant becomes $\lambda_{tr}$=1.3. The reflectivity corresponding
to fit 3 is plotted as a dashed blue curve in Fig.\ref{R_fit}a.

The central frequencies $x_k$ and partial electron-phonon constants
$\lambda_k$ are shown as symbols in Fig.\ref{a2f_fig} for fit 2
(circles) and fit 3 (squares) described above. The Eliashberg
function from \cite{Lortz2005} used in fit 1 is plotted as a solid
curve. Central frequencies of the phonons are larger than those
derived from specific heat and resistivity measurements
\cite{Lortz2005} (140 cm$^{-1}$ and 420 cm$^{-1}$) (Fig.
\ref{a2f_fig}).
%Our preliminary Raman experiments on the same sample
%showed a set of modes centered around 180 cm$^{-1}$, 650 cm$^{-1}$,
%780 cm$^{-1}$ and 1120 cm$^{-1}$ \cite{Teyssier2006}. The lowest and
%highest modes correspond reasonably well to the two modes of
%$\alpha^2_{tr}F(\Omega)$. It is worth noting that in MgB$_2$, the
%coupling comes predominantly from the Raman active E$_{2g}$ mode.

A more traditional way to extract $\alpha_{tr}^2F(\Omega)$ from optical
conductivity is to use directly the inversion
formula\cite{Marsiglio1998,Marsiglio1999}:
\begin{equation}\label{dinvert1}
\alpha_{tr}^2F(\omega)=\frac{1}{2\pi}\frac{\Omega_p^2}{4\pi}\frac{d^2}{d\omega^2}\Big(\omega
\mbox{Re}\frac{1}{\sigma(\omega)}\Big)
\end{equation}
One can also calculate the integrated coupling constant:
\begin{equation}\label{dinvert2}
\lambda_{tr,int}(\omega)=2\int_0^\omega\frac{d\Omega\alpha_{tr}^2F(\Omega)}{\Omega}
\end{equation}

\begin{figure}
\includegraphics[width=8cm]{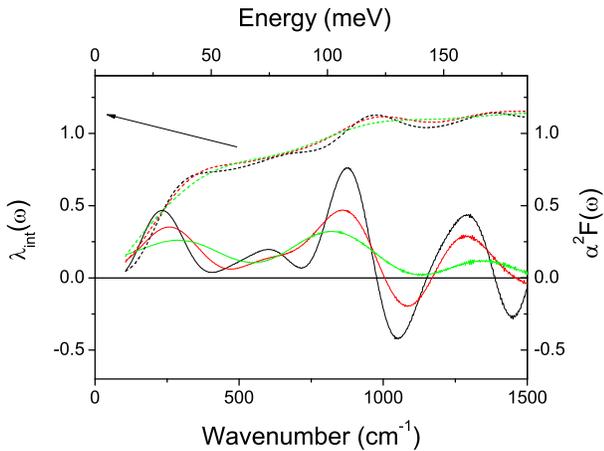}\\
\caption{(color online) $\alpha_{tr}^2F(\Omega)$ obtained using
Eq.(\ref{dinvert1}) at T=25 K (solid curves) and integrated coupling
constant $\lambda_{tr,int}(\omega)$ (dashed curves) for different
smoothening windows.}\label{a2F_frank}
\end{figure}

Taking the frequency derivative twice requires heavy spectral
smoothening of optical data. The result of inversion is given in
Fig.\ref{a2F_frank} for different frequency smoothening windows. The
robust features in $\alpha_{tr}^2F(\omega)$ are the peaks at about 250
cm$^{-1}$ and 900 cm$^{-1}$, in good agreement with the results
obtained by fitting procedures described above. Since
Eq.(\ref{dinvert1}) assumes the absence of low energy interband
transition, in our case some spurious features in
$\alpha_{tr}^2F(\omega)$ may appear at frequencies above $\sim$1000
cm$^{-1}$. Even though smoothening has a significant influence on
the shape of $\alpha_{tr}^2F(\omega)$, all curves $\lambda_{tr,int}(\omega)$
lie almost on top of each other and give a value of about 1.1 around
1000 cm$^{-1}$. Summarizing the results of all previously described approaches to extract $\alpha_{tr}^2F(\omega)$, we obtained $\lambda_{tr}=1.0\pm0.3$. This value indicates a medium to strong coupling in
ZrB$_{12}$.

\subsection{Temperature dependence of Drude spectral weight}
\label{sectionSW}

\begin{figure}[h]
\includegraphics[width=8cm]{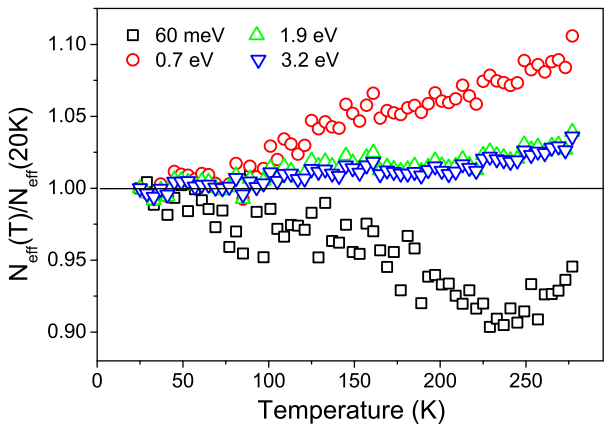}\\
\caption{(Color online) Temperature dependence of relative effective
number of carriers for different cutoff energies.}\label{SWT}
\end{figure}

The plasma frequency $\Omega_p$ at 25 K, obtained from different fits (Table
I) is 5.2-5.5 eV. The uncertainty is coming from the fact that interband contributions, which can not be separated unambiguously from the Drude peak. This value is slightly lower than the one calculated by LDA
(6.3 eV). However this agreement is good and it suggests that LDA
describes well the conduction bands near the Fermi surface. The
temperature dependence of the plasma frequency though appears to be
anomalous. As it was mentioned above, the integrated number of
carriers decreases for energies above 100 meV when temperature is
decreased. This is shown in Fig.\ref{SWT} where we plot $N_{eff}$ as
a function of temperature normalized by its value at 20 K for
selected cutoff frequencies. The effect is maximal (about 10\%) at
around 0.7 eV. Normally, the narrowing of the Drude peak with a
temperature independent plasma frequency results in a slight (1-2\%)
increase of $N_{eff}$ at frequencies 0.5-1 eV. In our case one has
to assume that the plasma frequency decreases with cooling down. On
the other hand, spectral weight is almost recovered at around 2 eV
which is due to the fact that it is transferred from the low
frequency range to the region between 0.5 and 2 eV
(Fig.\ref{S1temp}).

We do not have a definitive explanation of this effect. In order to
check the direct effect of temperature on the band structure, we
have performed LDA calculations at finite temperatures by adding a
Fermi-Dirac distribution for the bands filling in the self
consistent calculation. The thermal expansion was also taken into
account using data from Ref.\onlinecite{Lortz2005}. We found that
the calculated plasma frequency decreases around $0.4\%$ with
increasing temperature from 0 K to 600 K. Both sign and amplitude of
the change in the calculated plasma frequency cannot explain the
experimental observations. The spectral weight due to interband
transitions has also been calculated and found to be almost
temperature independent. Therefore a decrease of 10\% of the
spectral weight is so big that it likely results from a significant
change of the electronic band structure.

\begin{figure}
\includegraphics[width=8cm]{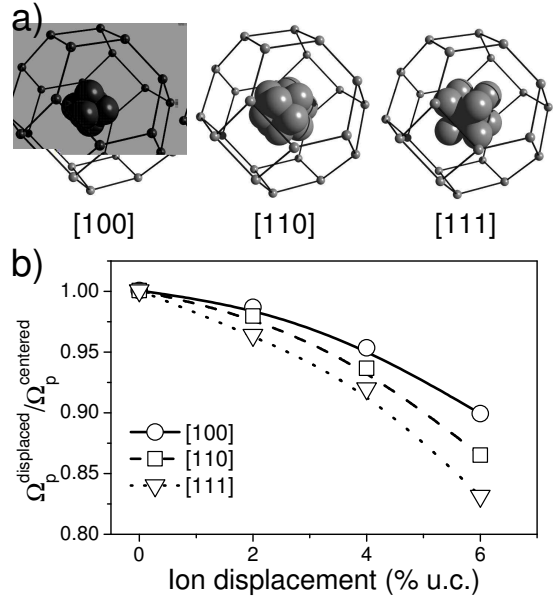}\\
\caption{a) Representation of displacement directions in the
$B_{24}$ cage. b) Plasma frequency versus ionic displacement along
[100], [110] and [111] directions.}\label{diplacement}
\end{figure}

One can speculate that a certain structural instability is due to
the ability of zirconium atoms to move almost freely in spacious
boron cages. According to Ref.\onlinecite{Werheit2005}, a
displacement of the metal ion in the $B_{24}$ cage of about $5\%$ of
the unit cell is responsible for the symmetry breaking which makes
low frequency Zr mode Raman-active. We have therefore calculated the
band structure and the plasma frequency as a function of
displacement of the metal ion in the directions [100],[110] and
[111]. The delocalization configurations are sketched in
Fig.\ref{diplacement}a.

Electron states at the Fermi surface are equally coming from Zr
atoms and $B_{12}$ cluster (Fig.\ref{bands}b). Therefore the highest
plasma frequency is obtained when Zr ion is in the center of the
B$_{24}$ cluster. The plasma frequency decrease will be more
pronounced when Zr shifts towards a region with lower charge density
(Fig.\ref{diplacement}b).

%The displacement configuration retainded for LuB$_{12}$ was along
%[100] direction with an occupancy factor for the delocalized site of
%$1.3\%$.

Such an anomalously large atomic displacement has already been
observed in clathrate-type systems that present a similar cluster
structure \cite{Chakoumakos2000,Chakoumakos2001}. In these
compounds, the displacement of an ion located in the middle of a
cage is strongly anisotropic and can be described by a fractionally
occupied split site. It was also found that this atomic disorder was
present from room temperature down to 10K.

In the picture of a multiple well potential, the delocalization due
to thermal excitation tends to center the ion in the middle of the
cage. At low temperature, the ion is located in the bottom of the
well out of the central position. Further structural investigations
are necessary to elucidate the relevancy of this or other scenarios in
dodecaborides. This kind of spectral weight transfer is also predicted by the model where the charge carriers are small polarons\cite{Emin1993}. While keeping in mind that they have been derived for a low charge carrier density, the observed direction of the spectral weight transfer agrees with those models.

\section{conclusions}

In this work, optical properties of zirconium dodecaboride have been
studied as a function of temperature. We found that LDA theory
predicts correctly interband optical conductivity and free electron
plasma frequency.

ZrB$_{12}$ appears to be a good metal with a bare plasma frequency
of about 5.5 eV. The Eliashberg function $\alpha^2_{tr}F(\Omega)$
was extrated by two methods giving consistent results. Two peaks
where found in $\alpha^2_{tr}F(\Omega)$ at about 200 and 1000
cm$^{-1}$ which is consistent with specific heat measurements
\cite{Lortz2005}. A medium to high electron-phonon coupling regime
is found with $\lambda_{tr}\simeq$1.0 in agreement with a high ratio
$2\Delta_o/kT_c=4.8$ from tunneling measurements
\cite{Tsindlekht2004}. The low frequency peak that likely
corresponds to a vibration mode of zirconium in boron cages is the
main contributor to the coupling constant.

However, this compound shows deviations from conventional metallic
behavior. Namely, a significant anomaly in the temperature evolution
of the integrated spectral weight was observed. About 10\% of the
spectral weight is removed from the Drude peak and transferred to
the region of low energy interband transitions. The origin of this
anomaly has not been clearly identified. However, our simulation of
the impact of the delocalization of the metal ion in the boron cages
shows that it could be responsible for the large change in the
plasma frequency.

\section{Acknowledgments}

The authors would like to thank Rolf Lortz and Alain Junod for
fruitful exchanges. This work was supported by the Swiss National
Science Foundation through the National Center of Competence in
Research Materials with Novel Electronic Properties-MaNEP.
FM greatly appreciates the hospitality of the Department of
Condensed Matter Physics at the University of Geneva. This work was
supported in part by the Natural Sciences and Engineering Research Council
of Canada (NSERC), by ICORE (Alberta), and by the Canadian Institute for
Advanced Research (CIAR).

%\bibliography{biblio}

\end{document}